\pdfoutput=1

\documentclass[aps,twocolumn,a4paper]{revtex4-1}
\usepackage[utf8]{inputenc}

\usepackage{amsmath}
\usepackage{amsfonts}
\usepackage{amssymb}
\usepackage{amsthm}
\usepackage{graphicx}
\usepackage[T1]{fontenc}
\usepackage{ae}
\usepackage{soul}  
\usepackage{bbold}  
\usepackage{hyperref}

\renewcommand{\vec}[1]{\mathbf{#1}}

\newcommand{\bra}[1]{\langle#1|}
\newcommand{\ket}[1]{|#1\rangle}

\newcommand{\up}{\uparrow}
\newcommand{\down}{\downarrow}

\newcommand{\Ttemp}{\mathrm{T}}

\newcommand{\Tsb} {\mathcal{T}_{SE}}
\newcommand{\Tns} {\mathcal{T}_{SN}}
\newcommand{\Tnssm}{\tilde{\mathcal{T}}_{SN}}
\newcommand{\Tr}{\mathrm{Tr}}

\begin{document}
\title{Spin-dependent thermoelectric transport in HgTe/CdTe quantum wells}
\author{ D. G. Rothe, E. M. Hankiewicz, B. Trauzettel, and M. Guigou}
\affiliation{Institute for Theoretical Physics and Astrophysics,
University of W$\ddot{u}$rzburg, 97074 W$\ddot{u}$rzburg, Germany}


\begin{abstract}
We analyze thermally induced spin and charge transport in HgTe/CdTe quantum wells
on the basis of the numerical non-equilibrium Green's function technique in the linear response regime.
In the topologically non-trivial regime, we find a clear signature of the gap of the edge
states due to their finite overlap from opposite sample boundaries -- both in the charge Seebeck and spin Nernst signal.
We are able to fully understand the physical origin of the thermoelectric transport signatures of edge and bulk states based on simple analytical models. Interestingly, we derive that the spin Nernst signal is related to the spin Hall conductance by a Mott-like relation which is exact to all orders in the temperature difference between the warm and the cold reservoir.
\end{abstract}

\maketitle

\section{Introduction}

Thermoelectric transport coefficients define the efficiency of a system to generate an electrical power from a temperature gradient \cite{MacDonald}.
The most established thermoelectric phenomenon is the Seebeck effect \cite{SeebeckA,SeebeckB}, in which a current (closed boundary conditions), or a bias (open boundary conditions) is induced from a temperature difference held between two reservoirs of a junction. The transverse Seebeck coefficient, or Nernst coefficient, refers to the alternative situation where the thermally induced current (bias) flows in the direction transverse to both the temperature gradient and the applied magnetic field \cite{Nernst}.

Thermoelectric effects have major consequences in terms of technological impact and scientific understanding.
On the one hand, these effects offer interesting applications based on heat-voltage conversion: thermometry, refrigeration, power generation \cite{Giazotto06,Bell}.
On the other hand, thermoelectric coefficients combine information from energy and charge flows at quasi-equilibrium. Furthermore, they are more sensitive to the details of the density of states than electrical conductance \cite{Abrikosov,Iiman,BeenakkerStaring}. Both aspects make them a powerful tool to probe the system dynamics \cite{Segal}.

During the last two decades, there have been considerable technological advances in low-temperature nanoscale physics. This allows precise measurements of thermoelectric transport signals, obtained in various systems like bismuth \cite{Behnia07}, superconductors \cite{Bel04,Chang10}, carbon-based structures \cite{Balandin}, or molecular junctions \cite{DubiDiVentra}.

The recent alliance of spintronics and thermoelectric transport brings up a spin analog of Seebeck and Nernst effects (see Ref.\ \cite{Bauer11} for a short review). Especially in systems with strong spin-orbit interactions, a temperature gradient can generate a transverse spin current (or bias) even in the absence of an applied magnetic field. This can lead to the anomalous Nernst effect (in the case of ferromagnetic systems)\cite{Goswami11,Miyasato07,Slachter11,Hanasaki08} or the spin Nernst effect (in the time-reversal symmetric situation)\cite{Chuu10,LiuXie10,Dyrdal11}.

Systems with strong spin-orbit interactions have been extensively studied in condensed matter physics especially since the prediction of the spin Hall effect 
\cite{Dyakonov2, Hirsch99, Murakami03, Sinova04}
 which allows for an all-electrical manipulation of spin. The spin Hall effect generates a transverse spin accumulation as a response to a longitudinal applied electric field. Spin-orbit interactions have several origins which distinguish the different types of phenomena, for instance, an \textit{extrinsic} spin Hall effect can emerge from the spin-orbit dependent scattering on impurities or defects \cite{Dyakonov2,Hirsch99,Hankiewicz06,Hankiewicz09,Kato,Garlid10}. On the other hand, bulk or structure inversion asymmetries 
 \cite{Rashba84, Dresselhaus55, Winkler} 
 give rise to an \textit{intrinsic} spin Hall effect \cite{Murakami03,Sinova04, Bruene10}, which may be described in terms of an anomalous velocity or a spin-dependent classical force \cite{Berry84,XiaoNiu10,SundaramNiu98}.

Recent experiments have demonstrated the existence of an intrinsic spin Hall effect in HgTe/CdTe quantum wells (QWs) \cite{Bruene12} by the use of the quantum spin Hall effect as the detector. This novel electronic phase is characterized by an insulating bulk and protected metallic edge states. 
The emergence of the quantum spin Hall effect is due to strong spin-orbit coupling and other relativistic corrections, 
which reverse bands of opposite parities. The electrons obey a massive Dirac equation and the sign of the mass term enables us to distinguish the topological phases. The edge states consist of Kramers pairs moving in opposite direction at each boundary \cite{Kane05QSHE,Fu07,Bernevig06} and time-reversal symmetry protects them from non-magnetic and elastic backscattering \cite{Wu06}. Thereby, these edge channels carry ``dissipation-less'' spin currents whose existence in HgTe QWs has been confirmed experimentally by measurements in multi-terminal devices \cite{Koenig08}.

Recently, topological insulators have been proposed as good materials for thermoelectric conversion \cite{Takahashi10,Tretiakov10,Tretiakov11,Murakami11}. The basic idea relies on the topological protection of 1D edge states which prevents reduction of electrical transport in disordered systems. The authors consider narrow ribbons of quantum spin Hall insulators or 3D topological insulator with line dislocations. The aim is the enhancement of the contribution of edge states to the thermoelectric transport compared to the bulk modes. Hence, the analysis is restricted to a small energy range excluding the valence bands. Inelastic processes are taken into account through the Boltzmann transport theory. The latter is used to calculate the thermoelectric coefficients in a two-band model. Thus, the efficiency of these systems to convert heat into electricity is based on the dominance of the edge modes on transport.

In this work, we investigate the spin-dependent thermoelectric transport in quantum spin Hall insulators based on HgTe/CdTe QWs in absence of magnetic fields. The behavior of the Seebeck coefficient and the spin Nernst signal is analyzed in a four-terminal cross-bar setup, as shown in Fig. \ref{drawing1}. A thermal gradient between lateral leads induces a longitudinal electrical bias and a transverse spin current. Each of them can be used as a probe of the topological regime as well as finite size effects of the quantum spin Hall insulator. We show that the oscillatory character of the Seebeck and spin Nernst coefficients in the bulk gap highlights the presence of the mini-gap -- due to the finite overlap of the edge states from opposite sample boundaries. Furthermore, we describe a qualitative relation between the type of particles in a given band and the magnitude of the spin Nernst signal. This allows us to provide a natural explanation of the observed phenomena based on anomalous velocities and spin-dependent scattering off sample boundaries.

The article is organized as follows. In Sec. II, we introduce the model Hamiltonian of the HgTe/CdTe QW and describe the formalism necessary to calculate the Seebeck and spin Nernst coefficients. The thermoelectric transport by the edge states -- with a particular emphasis on finite size effects -- is analyzed in Sec. III through the behavior of Seebeck and spin Nernst signals. In Sec. IV, we focus on the spin-dependent thermoelectric effect induced by the bulk states. We conclude in Sec. V and present details of the calculation in the appendices.

\section{Model}

In this section, we present the model Hamiltonian of the HgTe/CdTe QW and give the general expressions of the Seebeck and spin Nernst coefficients.

\subsection{Hamiltonian}

We consider a four-terminal cross-bar setup based on a HgTe/CdTe QW whose low-energy dynamics is described by the Bernevig-Hughes-Zhang (BHZ) 4-band model \cite{Bernevig06,Koenig08}.
The  Hamiltonian is written in the basis of the lowest QW subbands
$\ket{E+}$, $\ket{H+}$, $\ket{E-}$,  and $\ket{H-}$. Here, $\pm$ stands for two Kramers partners but in the following, we will refer to them as spin components, denoted by $\up$,$\down$, for brevity. The spin z-direction corresponds to the QW growth direction, which is [001]. The Hamiltonian can be written as
\begin{align}
\label{H4bdmodel}
 H = V_m(\vec{r})\tau_z - D k^2  + \left( \begin{array}{cc} h(k) & 0 \\
                                 0   & h^*(-k)
               \end{array} \right),
\end{align}
\begin{align}
h(k) = \left( \begin{array}{cc} \mathcal{M}(k) & A k_+ \\
                                 A k_-   & -\mathcal{M}(k)
               \end{array} \right)
\end{align}
with $k^2 = k_x^2 + k_y^2$, $k_\pm = k_x \pm i k_y$, and $\mathcal{M}(k) = M - B k^2$.
The sign of the gap parameter $M$ determines whether we are in the regime
of a trivial insulator ($M > 0$) or a topological insulator ($M<0$). Experimentally, $M$ is
tuned by changing the QW width.

The term $V_m(\vec{r}) \tau_z$ describes an in-plane confinement potential, where $\tau_z$ is a Pauli matrix acting on the $E/H$ space. By this kind of confinement we may ensure
that outside of the sample, i.e. in vacuum, the parameter regime is topologically trivial, so that edge states, if present, will
be confined. Calling the inside of the sample $G$, the limit $V_m(\vec{r}) \to \infty \; \forall  \vec{r} \in \partial G$  can be used to
make all components of $\psi$ vanish at the sample boundary, the envelope function $\psi$ being
the solution of the Dirac equation based on the Hamiltonian (\ref{H4bdmodel}).

We mention in passing that this model can be extended by a term breaking the structural inversion asymmetry (SIA) with a z-dependent potential. The resulting Rashba-like interaction connects the Kramers blocks of the Hamiltonian (\ref{H4bdmodel}) affecting the particles with opposite spin
\begin{align} \label{hra}
h_R(k) = \left( \begin{array}{cc} -iR_0 k_- & -i S_0 k_-^2 \\
                                 i S_0 k_-^2 & i T_0 k_-^3
               \end{array} \right)
\end{align} 
with the Rashba coupling parameters $R_0$, $S_0$, and $T_0$ \cite{Rothe10}. We have analyzed that such a term will only quantitatively affect all our results presented below. Therefore, we will not further consider effects due to SIA in this article.

\begin{figure}
\includegraphics[scale=.32]{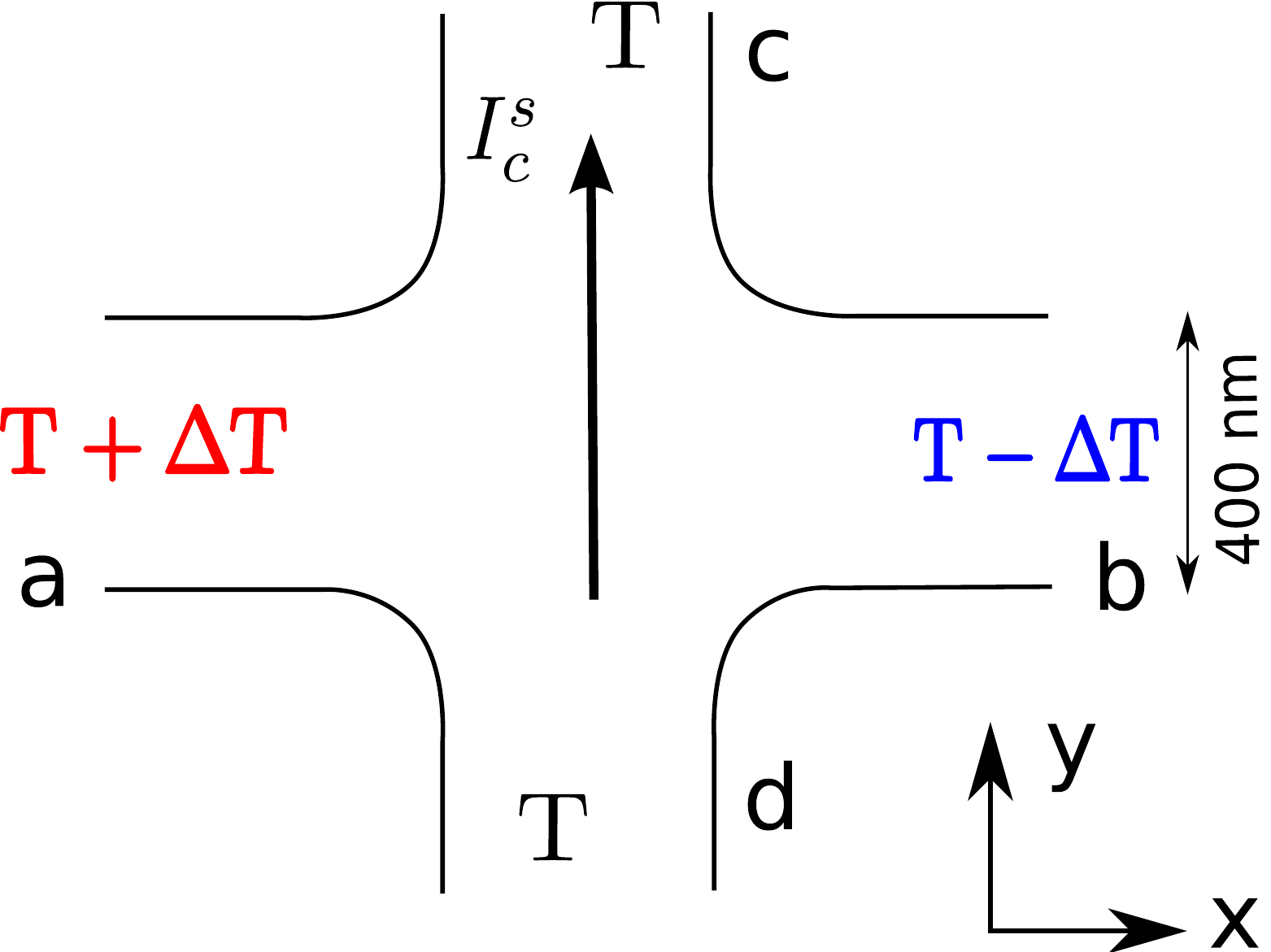}
\caption{\label{drawing1}
(Color online) Four-terminal cross-bar setup based on HgTe/CdTe QWs used for thermoelectric transport.
A longitudinal temperature gradient $\Delta \Ttemp$ is applied between reservoirs $\mathsf{a}$ and $\mathsf{b}$ and generates a transverse spin current
$I_{\mathsf{c}}^s$ detected, for instance, in reservoir $\mathsf{c}$.
}
\end{figure}

Figure \ref{drawing1} shows the four-terminal cross-bar setup we analyze. The central sample is connected to four semi-infinite leads: the reservoirs $\mathsf{a}$ and $\mathsf{b}$ are maintained respectively at warmer and colder temperature than the rest of the system -- creating a longitudinal temperature gradient -- while the transverse terminals $\mathsf{c}$ and $\mathsf{d}$ are used to probe spin currents.
\begin{figure}
(a)\includegraphics[scale=.6]{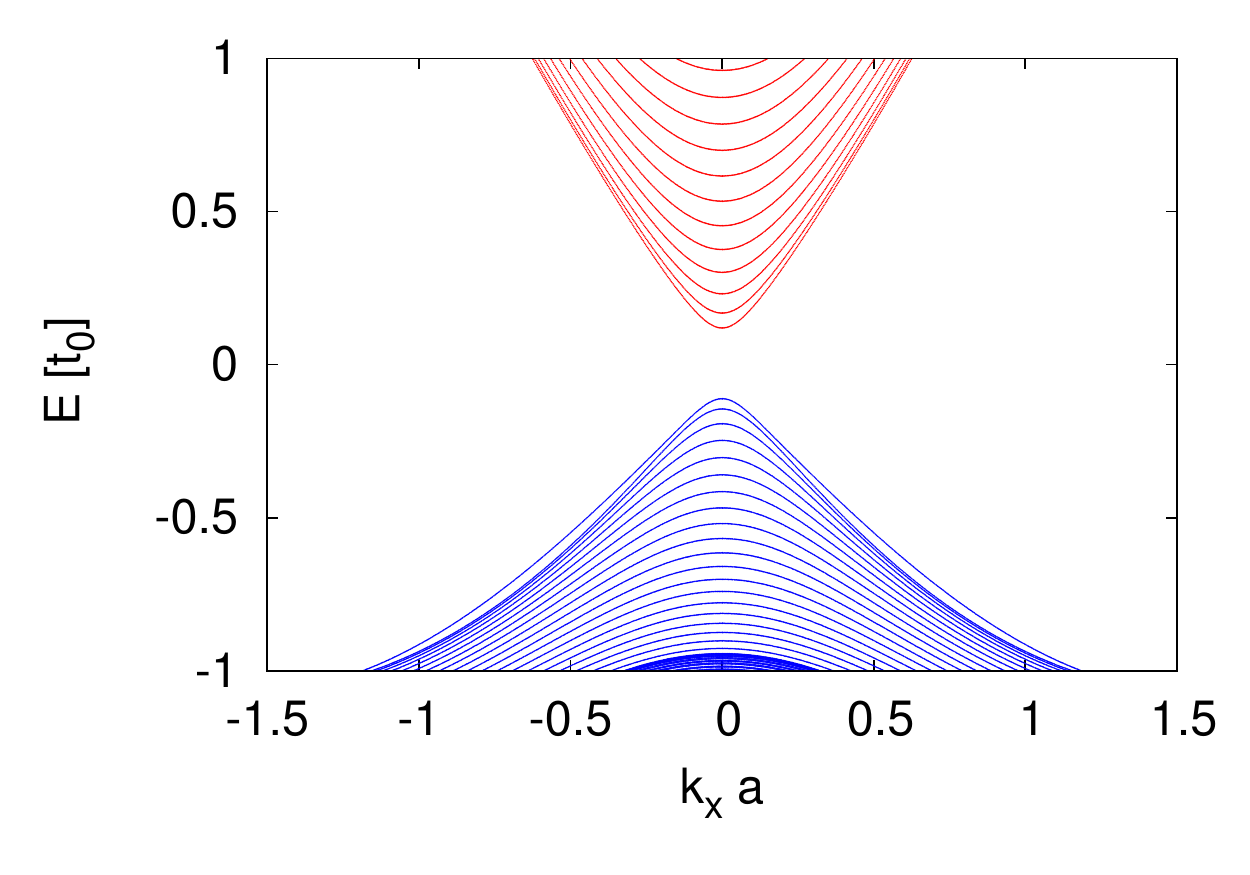}
\\
(b)\includegraphics[scale=.6]{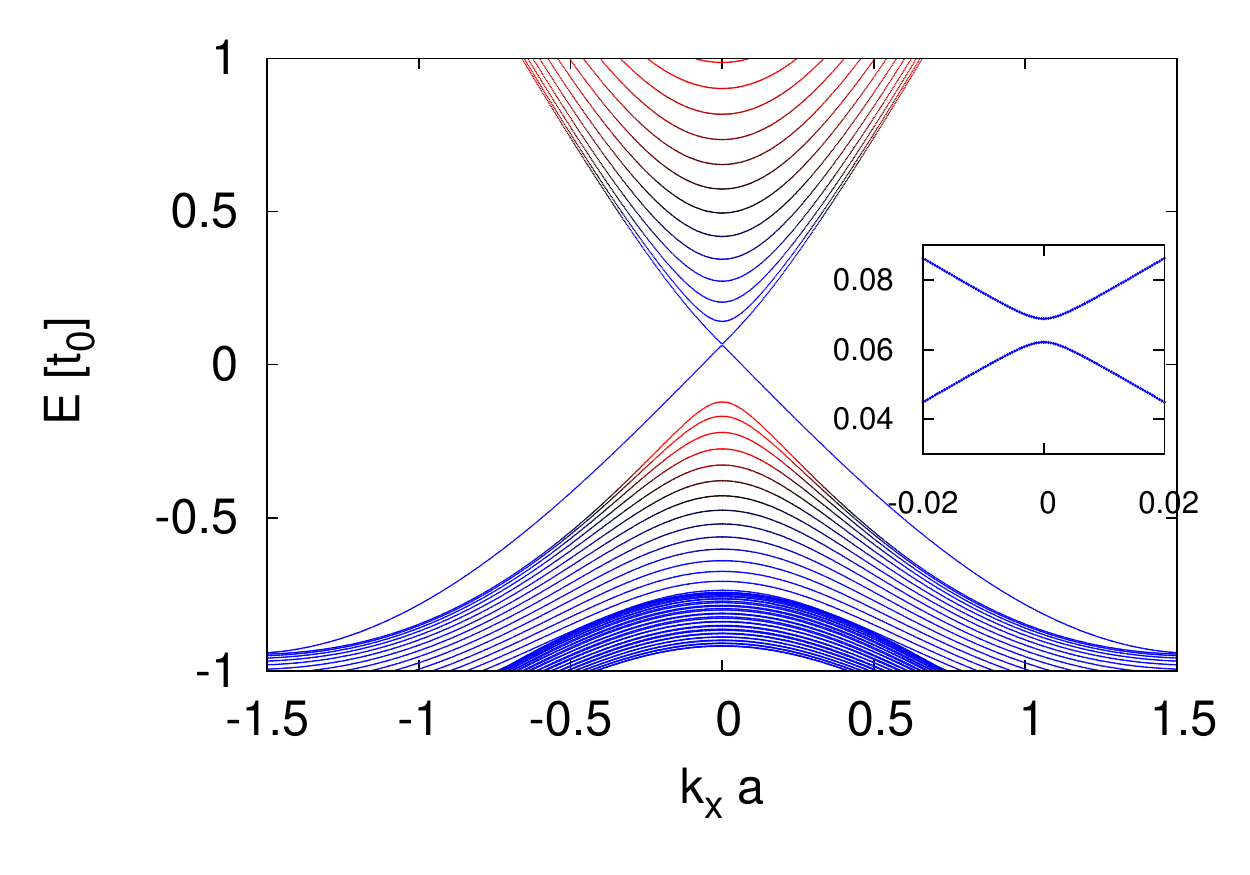}
\caption{\label{dispNINV}
(Color online) Subband dispersion relation for the leads (400 nm wide), in
(a) the normal regime $(M = 0.1 t_0)$ and
(b) the inverted regime $(M = -0.1 t_0)$. The inset shows the mini-gap in the dispersion caused by the overlap of edge states.
The coloring highlights the transition from electron-like character ($\ket{E\pm}$ in red) to heavy-hole character
($\ket{H\pm}$ in blue)
with full color for a 20\% excess of either contribution.
The lattice constant is $a = 6.6 \mathrm{~nm}$ so $t_0 = \mathrm{44~meV}$. The mini-gap width
is $6.5 \times10^{-3} ~ t_0( = 0.28 \mathrm{~meV})$.
}
\end{figure}

To model the setup and treat the thermoelectric transport properties, we employ the tight binding approach.
Therefore, we discretize the continuum model \eqref{H4bdmodel} on a lattice of spacing $a$
 by the substitutions $k_i^2 \to \frac{1}{a^2} ( 2 - 2 \cos k_i a)$ and $k_i \to \frac{1}{a} \sin ( k_i a) $, where $i$ is the index of the lattice site.
The confinement potential is implemented by the lattice truncation in accordance with the geometry of the sample.
Rewriting the trigonometric functions in terms of translation operators, this leads to a tight binding
 Hamiltonian which only contains nearest-neighbor hopping terms between the lattice sites (see Appendix \ref{appendixTBModel} for details).
The energies of the model are expressed as functions of the conduction band hopping parameter $t_0 = -\frac{D+B}{2 a^2}$ where the parameter values
 of the HgTe/CdTe QW are taken as in typical experiments, \textit{i.e.} $A = 0.375 \mathrm{~eV nm}$, $B = -1.120 \mathrm{~eV nm^2}$, and $D = -0.730 \mathrm{~eV nm^2}$.
In the low-energy regime, the lattice constant is set sufficiently small compared to the Fermi wave length. Hence, for $a=6.6\mathrm{~nm}$,
the energy unit is $t_0=44\mathrm{~meV}$.
The parameter $M$ is chosen as $|M| = 0.1 t_0 = 4.4 \mathrm{~meV}$.
In Fig. \ref{dispNINV}(a) and (b), we show the subband dispersion relation for a HgTe/CdTe QW of width 400 nm,
both in the normal insulator and the topological insulator regimes. In the latter case, finite size effects emerge on the edge states since they substantially overlap \cite{Zhou08}. One of the consequences of this is the opening of a mini-gap, as shown in the inset of Fig. \ref{dispNINV}(b).

\subsection{Landauer B\"uttiker formalism and thermoelectric coefficients}

The particle current $I_{p\sigma}$ in the lead $p$ with spin $\sigma$ is obtained by the Landauer-B{\"u}ttiker formula \cite{Datta}
\begin{align}
I_{p\sigma} = \frac{1}{h} \sum_{q \neq p}\int dE \,  T_{p\sigma,q}(E) (f_p  - f_q )
\label{PartCurrent}
\end{align}
with $f_p = (e^{(E-\mu_p)/k_B \Ttemp_p} + 1)^{-1}$ the electronic Fermi distribution function, $k_B$ the Boltzmann constant, $\Ttemp_p$ the temperature, and $\mu_p$ the chemical potential.
The transmission probability $T_{p\sigma,q}(E)=\sum_{\sigma'}T_{p\sigma,q\sigma'}(E)$ from lead $p$ with spin $\sigma$ to lead $q$ can be evaluated using the non-equilibrium Green's function formalism (NEGF)  \cite{MeirWingreen, Sanvito99, WimmerPHD}
\begin{align}
T_{p\sigma,q\sigma'}(E) = \mathrm{Tr} [ \Gamma_{p\sigma} G^R \Gamma_{q\sigma'} G^A ],  \quad (p,\sigma) \ne (q,\sigma'),
\label{Trans_Proba}
\end{align}
where $\Gamma_{p\sigma}(E) = i( \Sigma_{p\sigma} - \Sigma_{p\sigma}^\dagger)$ refers to projectors on velocity operators of the propagating modes,
and $\Sigma_{p\sigma}(E)$ stands for the spin-dependent self-energy. The latter is defined by $\Sigma_{p\sigma}(E) = \tau_{p\sigma} (E + i 0^+ - H_{\rm leads})^{-1} \tau_{p\sigma}^\dagger$, where the matrix $\tau_{p\sigma}$ connects the lead $p$, spin $\sigma$ to the adjacent sites of the sample.
$G^R(E)=(G^A(E))^\dagger = (E - H_{\rm sample} - \sum_{p,\sigma} \Sigma_{p\sigma})^{-1}$ is the retarded Green's function.
Further, $H_{\rm leads}$ and $H_{\rm sample}$ represent, respectively, the lattice Hamiltonians of the decoupled leads and the sample.
Once the transmission probabilities are evaluated, the charge current $I^e_p=e(I_{p\uparrow}+I_{p\downarrow})$ and the spin current $I^s_p=(\hbar/2)(I_{p\uparrow}-I_{p\downarrow})$ can be obtained.

We consider a longitudinal temperature gradient between the leads $\mathsf{a}$ and $\mathsf{b}$ by setting $\Ttemp_{\mathsf{a}} = \Ttemp + \Delta \Ttemp$,
$\Ttemp_{\mathsf{b}} = \Ttemp - \Delta \Ttemp$, and $\Ttemp_{\mathsf{c}} = \Ttemp_{\mathsf{d}}  = \Ttemp$.
In the linear response regime, the Seebeck coefficient reports the longitudinal voltage bias $\Delta \mu=\frac{\mu_{\mathsf{a}}-\mu_{\mathsf{b}}}{2}$ generated by the temperature gradient $\Delta \Ttemp$ under the condition of vanishing charge currents (open boundary conditions).
Upon Taylor expansion of the Fermi functions in $\Delta \Ttemp$ and $\Delta \mu$, the Seebeck coefficient can be written as
\begin{align}
\label{SeebInt}
S_e  = \left. \frac{-\Delta \mu}{e \Delta \Ttemp} \right|_{I_{\mathsf{a},\mathsf{b}}^e=0}  \approx \frac{1}{e \Ttemp} \frac{ \int dE \, f_0 (1-f_0) \Tsb(E) (E-\mu)} {\int dE \, f_0 (1-f_0) \Tsb(E)}
\end{align}
with $f_0 =  (e^{(E-\mu)/k_B \Ttemp} + 1)^{-1}$ the Fermi distribution function at equilibrium.
In the above equation, we defined the Seebeck transmission function $\Tsb(E) = T_{\mathsf{a},\mathsf{b}} + (T_{\mathsf{a},\mathsf{c}} + T_{\mathsf{a},\mathsf{d}})/2$, where the summation over spins is implied.

Due to the presence of intrinsic spin-orbit interaction in the sample, the longitudinal thermal gradient $\Delta \Ttemp$ also yields a transverse spin current $I_{\mathsf{c}}^s(=-I_{\mathsf{d}}^s)$ in the case of closed boundary conditions.
 The spin Nernst coefficient is then defined as the ratio
\begin{eqnarray}
\label{NScoeff}
N_s &=& \left.\frac{I_c^s}{2\Delta \Ttemp}\right|_{\mu_{\mathsf{c},\mathsf{d}}=\mu} \nonumber\\
&\approx&  \frac{1}{8 \pi k_B \Ttemp^2} \int dE \, f_0 (1-f_0) \Tns(E) (E-\mu).
\end{eqnarray}
Here, we introduced the spin Nernst transmission function
$\Tns(E)  = \Delta T_{\mathsf{c},\mathsf{b}} - \Delta T_{\mathsf{c},\mathsf{a}}$,
with the short-hand notation
$\Delta T_{\mathsf{c},\mathsf{b}}(E) = T_{\mathsf{c}\up,\mathsf{b}\up}+T_{\mathsf{c}\up,\mathsf{b}\down}-T_{\mathsf{c}\down,\mathsf{b}\up}-T_{\mathsf{c}\down,\mathsf{b}\down}$.

Interestingly, the Mott relation provides information about the (spin-)thermotransport coefficients on the basis of the energy dependence of the (spin-)conductance \cite{CutlerMott69}.
In the low temperature limit, one can derive
\begin{align}
\label{Mott0}
S_e \approx \frac{\pi^2 k_B^2 \Ttemp}{3 e} \left.\frac{d \ln G_{xx}(E)}{dE}\right|_{E=\mu},
\end{align}
where $G_{xx}(E,\Ttemp=0)=\frac{e^2}{h} \Tsb(E)$ is the longitudinal conductance for zero temperature. Equation \eqref{Mott0} is valid 
if $k_B \Ttemp$ is large compared to the scale on which $\Tsb(E)$ varies. 
A numerical analysis in \cite{Lunde05} claims that \eqref{Mott0} can be valid even if $\Tsb(E)$ varies more rapidly, as long as 
$k_B \Ttemp \ll \mu$.

An analogous relation exists between the spin Nernst signal and the spin Hall conductance. From the Sommerfeld expansion of the transmission function in Eq. (\ref{NScoeff}), one obtains the following Mott-like formula
\begin{align}
\label{Mottlike}
N_s \approx \frac{ 2\pi^2 k_B^2 \Ttemp}{3 e} \left.\frac{d G_{sH}(E)}{dE}\right|_{E=\mu(\Ttemp=0)}
\end{align}
with $G_{sH}(E,\Ttemp=0)=\frac{e}{8\pi}\Tns(E)$, the spin Hall conductance at zero temperature.
In Appendix \ref{appendixMottlike}, we demonstrate that this relation can be extended to finite temperature
by defining a smoothed function $\Tnssm (E)$ (see Eq. \eqref{B5}) that depends on the temperatures in the leads.
As a result, we find an exact Mott-like formula for the spin Nernst coefficient
\begin{equation}
\label{MottlikeExactPaper}
 N_s(\mu)
 = \frac{\pi k_B^2 \Ttemp}{12} \left.\frac{d \Tnssm(E)}{d E}\right|_{E=\mu}.
\end{equation}
Since $\Tns(E)$ shows a highly oscillating behavior, the above equation simplifies the interpretation of the spin Nernst signal in terms of transmission functions because of the smoothing of $\Tnssm (E)$. Equation (\ref{MottlikeExactPaper}) is one of the key results of our paper.

\section{Thermoelectric transport carried by the edge states}

In this section, we present the numerical results of the spin Nernst and Seebeck coefficients for an energy regime within the bulk gap. When the HgTe/CdTe QW is in a topologically trivial phase, there is no sub-gap transport through the system. The transmission functions $\Tsb$ and $\Tns$ are then zero, and from Eqs. (\ref{SeebInt}) and (\ref{NScoeff}), it follows that there are no thermoelectric signals. On the contrary, the HgTe/CdTe QW in a non trivial phase hosts edge states in the bulk insulating gap. These modes carry electrons with opposite spins in opposite directions.

\begin{figure}
\includegraphics[scale=.3]{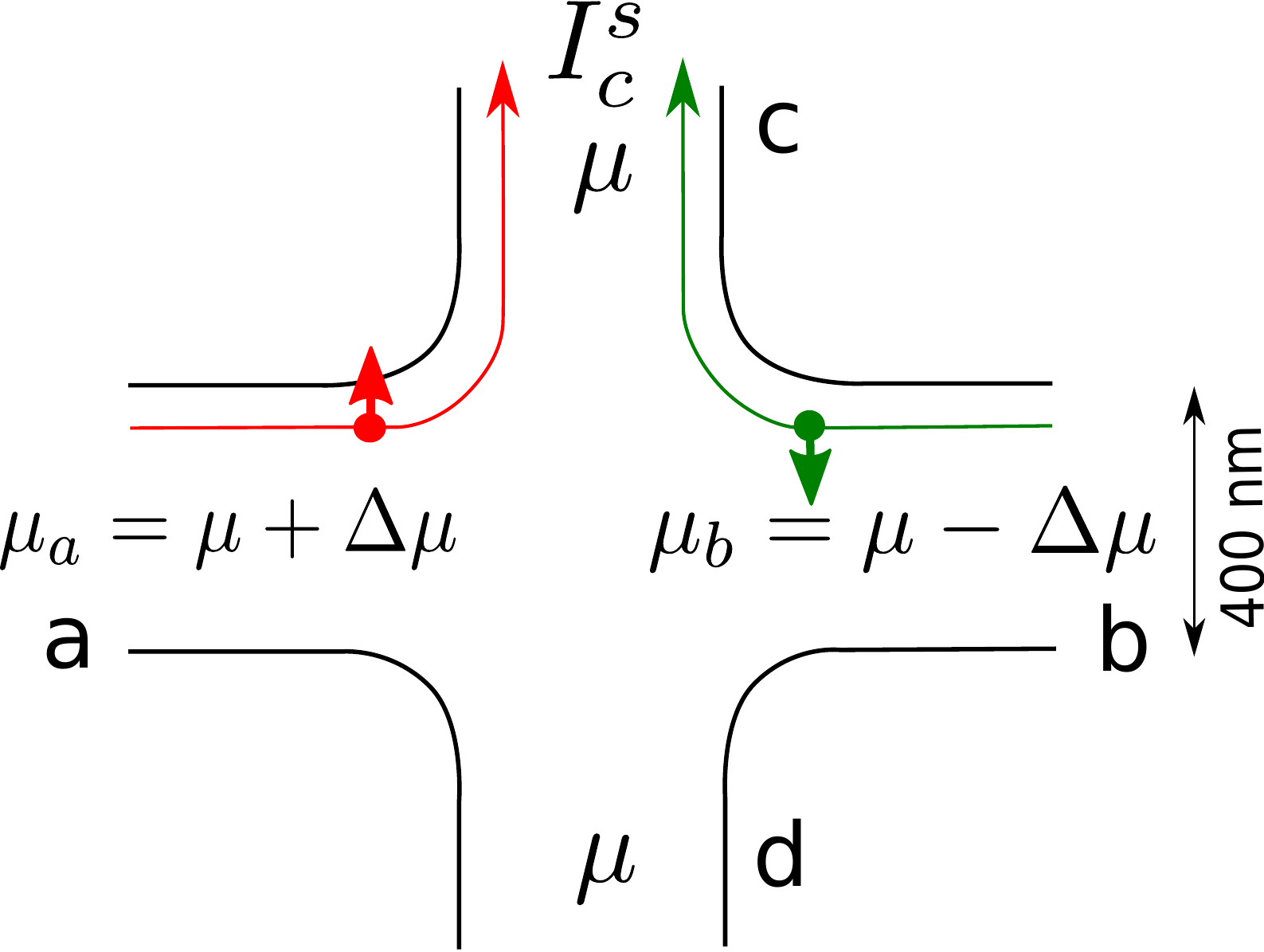} 
\caption{
\label{sampleEdgestates}
(Color online) Four-terminal cross-bar setup based on a HgTe/CdTe QW in the inverted regime. When an electrical bias is imposed between the longitudinal leads $\mathsf{a}$ and $\mathsf{b}$, one edge channel carries electrons with spin up from lead $\mathsf{a}$ to lead $\mathsf{c}$ (red solid line) and one edge channel carries electrons with spin down from lead $\mathsf{b}$ to lead $\mathsf{c}$ (green solid line). This gives rise to a spin current in lead $\mathsf{c}$ and, at zero temperature, to a quantized spin Hall conductance.
}
\end{figure}

With respect to the geometry of the setup, spin and electrical currents are induced and flow respectively in transverse and longitudinal leads, as depicted in Fig. \ref{sampleEdgestates}. However, the finite width of the system implies an overlap of the edge states meaning that backscattering processes can occur.

\begin{figure}
(a)\includegraphics[scale=.33]{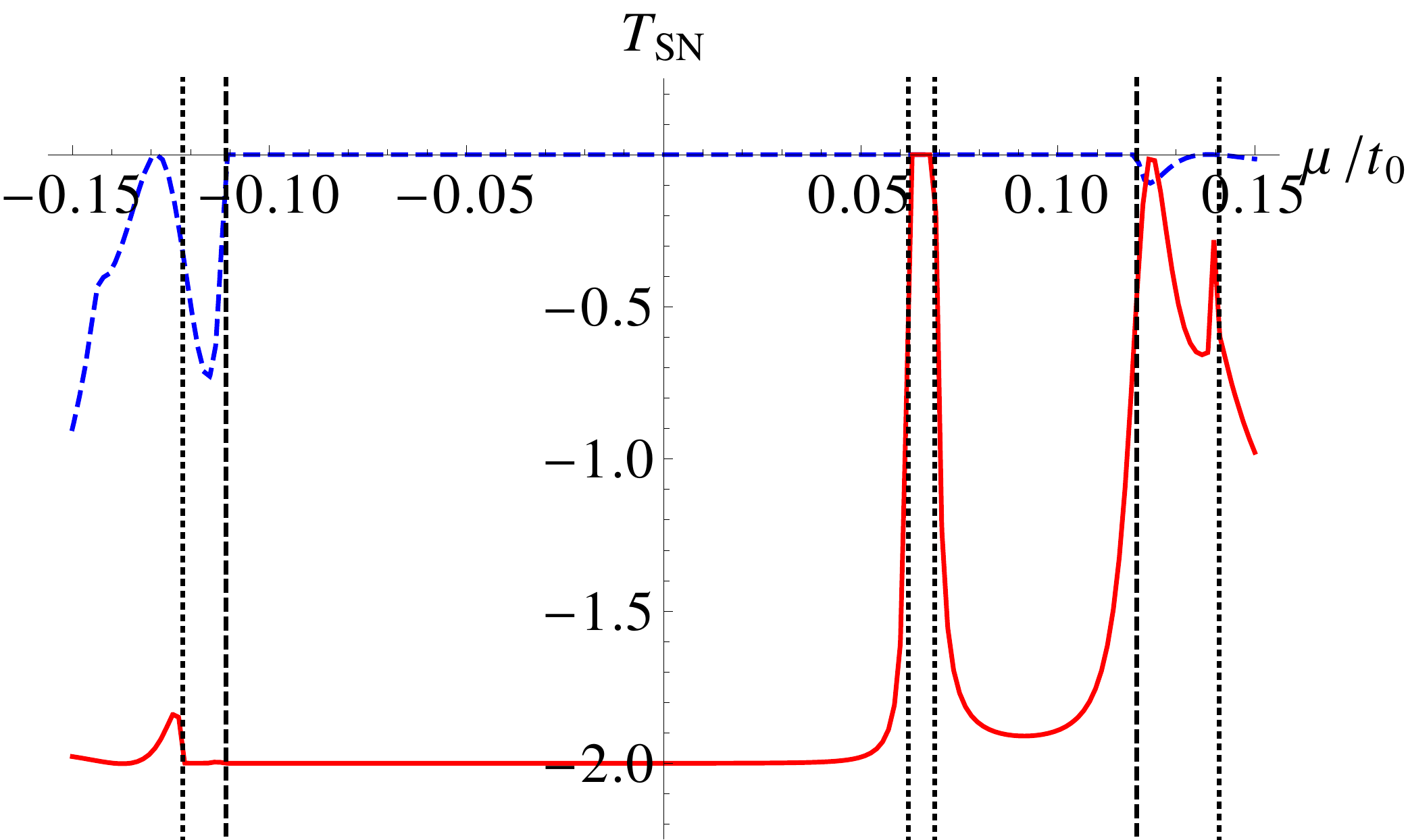}
\\
(b)\includegraphics[scale=.33]{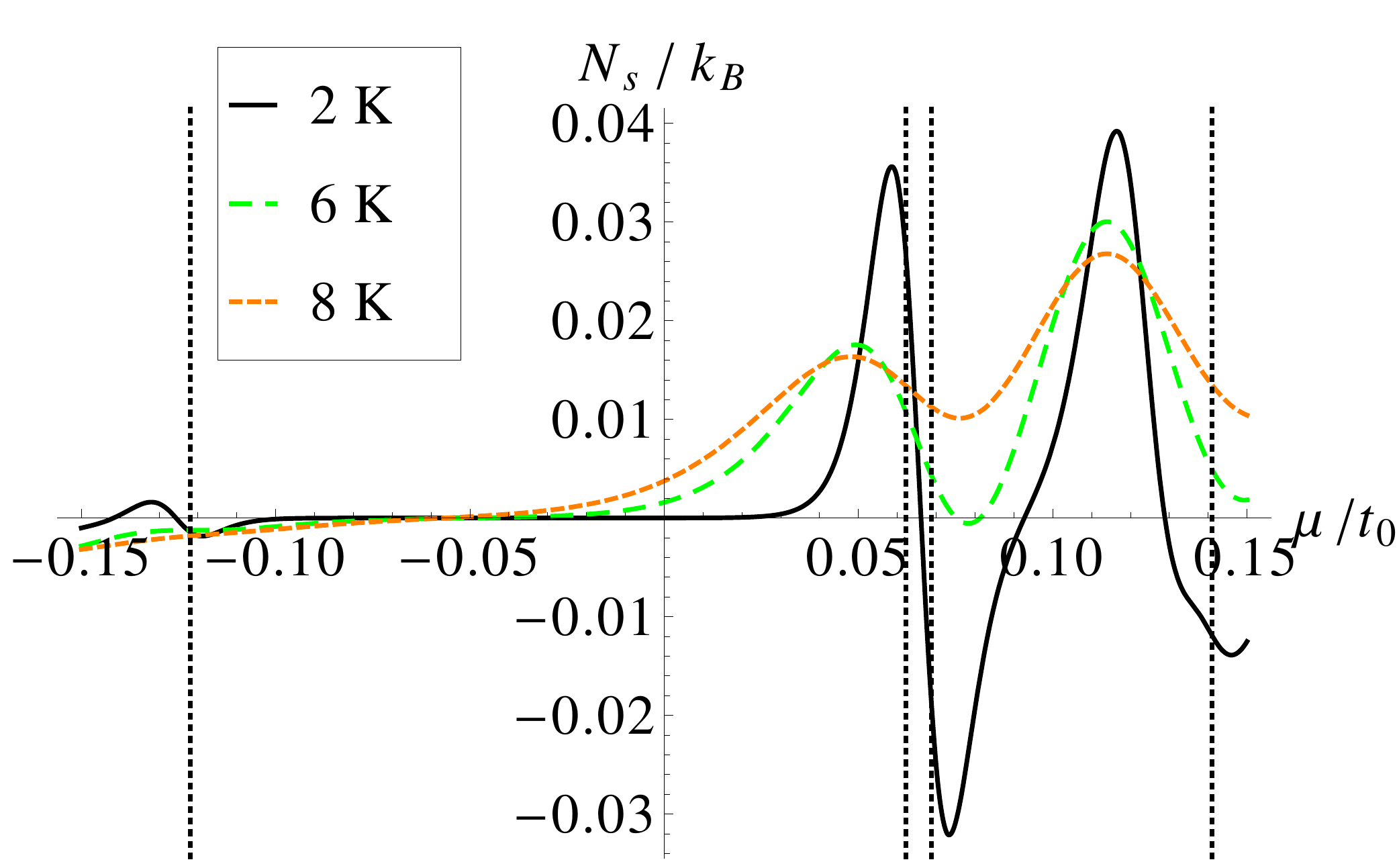}
\caption{
\label{snernstGap}
(Color online) (a)
Spin Nernst transmission function as a function of the chemical potential when the sample is in the normal (dashed blue line) or in the inverted (solid red line) regime.
The dotted vertical lines indicate the bulk gap and minigap positions in a finite system for the inverted regime, while the dashed vertical lines indicate the
gap in the normal regime.
(b) Spin Nernst signal $N_s/k_B$ in a system in the inverted regime at $\Ttemp= \mathrm{2~K}$ (black solid line), $\Ttemp= \mathrm{6~K}$ (green dashed line), and $\Ttemp= \mathrm{8~K}$ (orange dotted line).
}
\end{figure}

We first investigate the behavior of the spin Nernst signal $N_s$ and the associated transmission function $\Tns$. The results are presented in Figs. \ref{snernstGap}(a) and (b).
While the chemical potential is in the bulk gap, the spin transport is mediated by the edge channels so that the transmission function is simply given by
\begin{eqnarray}
\Tns = (T_{\mathsf{c}\up,\mathsf{b}}-T_{\mathsf{c}\down,\mathsf{b}}) - (T_{\mathsf{c}\up,\mathsf{a}} - T_{\mathsf{c}\down,\mathsf{a}}) = -2.
\end{eqnarray}
Evidently, as the chemical potential reaches the boundary of the mini-gap, the number of propagating states drops to zero and transport breaks down. This results in a peak of the transmission function $\Tns$. Consequently, the spin Nernst coefficient is zero in the bulk gap except when the chemical potential reaches the boundary of the mini-gap. 
Because of the Mott-like relation \eqref{MottlikeExactPaper},
a symmetric function $\Tns(E)$ must result in an antisymmetric function $N_s(E)$.
Therefore, $N_s$ exhibits an approximately antisymmetric peak centered at the maximum of the transmission peak.
The confinement of the QW implies an energy shift in the band dispersion. Therefore, the boundaries of the bulk gap are not exactly at energy $|M|=0.1 t_0$, as we can see
 in Figs. \ref{dispNINV} and \ref{snernstGap}.
In Fig. \ref{snernstGap}, the gap and minigap positions of a finite system are indicated by vertical lines. Dotted vertical lines are used for 
the inverted regime and dashed lines for the normal regime. Interestingly, one observes that the merging of the edge state to the conduction band causes $\Tns$ to vanish 
already before the first bulk mode appears. Where $\mu$ lies between the right dashed and dotted vertical lines, a finite $\Tns$ reappears due to the formation of the first 
bulk state, in the same subband as the edge state.
Outside the gap indicated by the vertical lines, bulk states start to participate to the spin transport resulting in additional oscillations in $\Tns$ as a function of $\mu$. They transform into peaks of the spin Nernst coefficient whose existence is understood with the same arguments as for the mini-gap peak. Especially at positive chemical potential, the magnitude of the peak is comparable to that of the mini-gap and allows to mark the position where the edge states merge.

In Fig. \ref{snernstGap}(b), we show the behavior of the spin Nernst coefficient for different temperatures. As $k_B \Ttemp$ increases, the position of the peaks is slightly shifted to lower energy. The magnitude tends to decrease and the peak width is broadened. Up to $\Ttemp = \mathrm{6~K}$, the spin Nernst signal goes to zero between the peak that specify the position of the mini-gap and the edge state merging peak. Beyond this temperature, $N_s$ is smoothed out, so that it can not probe the edge state signal.

\begin{figure}
\includegraphics[scale=.3]{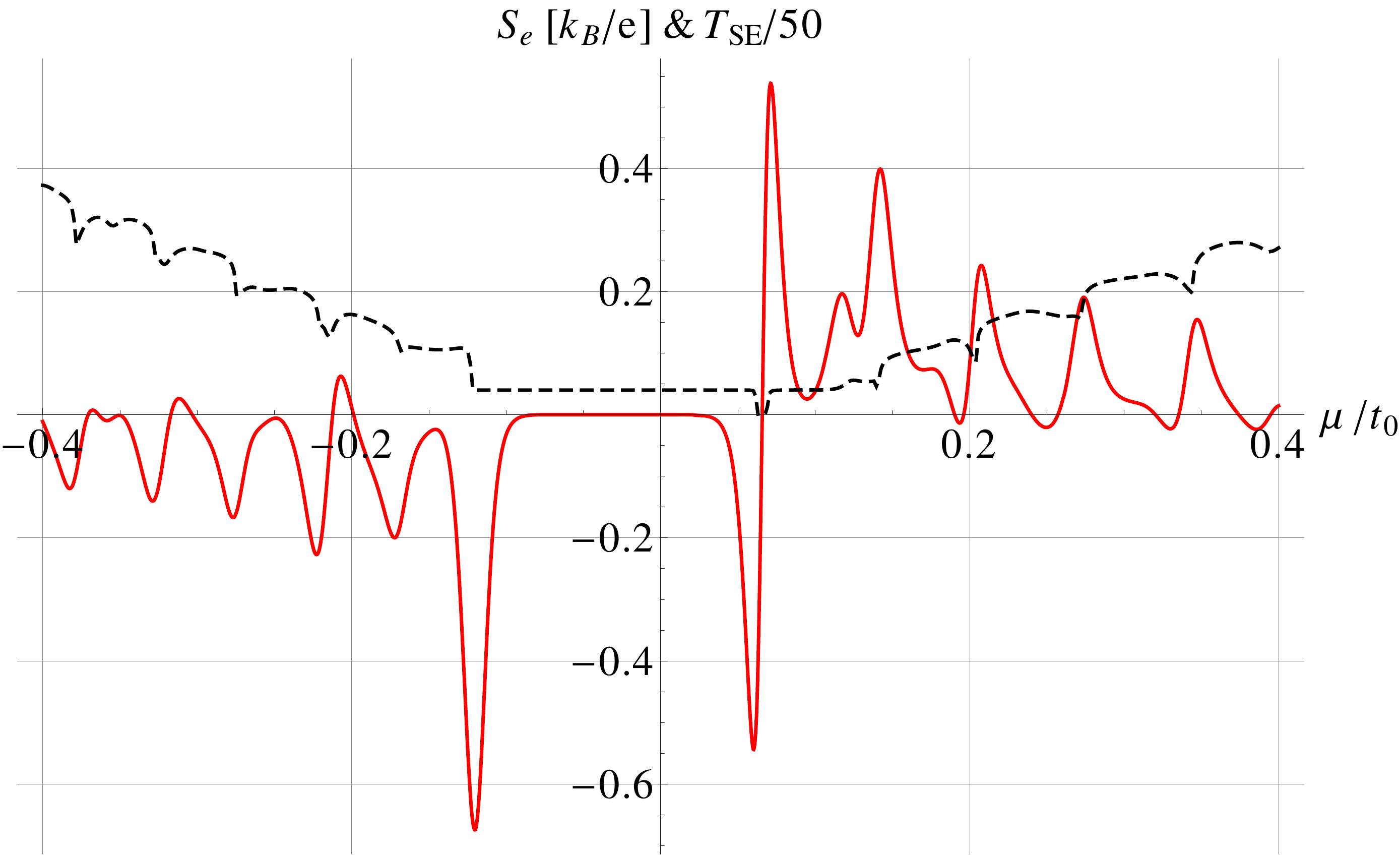}
\caption{\label{seebeck4inv}
(Color online) Seebeck coefficient $S_e [k_B/e]$ (red solid line) and scaled transmission function $\Tsb/50$ (black dashed line) as a function of the chemical potential at $\Ttemp = \mathrm{2~K}$. The mini-gap appears as an antisymmetric peak.
}
\end{figure}

We now turn to the analysis of the transmission function $\Tsb$ and the Seebeck coefficient $S_e$ as a function of energy. The results are presented in Fig. \ref{seebeck4inv}. Inside the bulk gap, the transmission function $\Tsb$ is constant but goes to zero when the chemical potential is in the mini-gap. This feature leads to an approximately antisymmetric peak in the behavior of $S_e$, which provides information on the presence and the position of the mini-gap in the spectrum. The boundary of the bulk gap manifests itself as the step of the transmission function and transforms as a narrow peak in $S_e$.

The transmission function $\Tsb$ exhibits a smoothed staircase behavior whose steps coincide with the opening (at positive energy) or the closing (at negative energy) of conducting channels. This behavior transforms into a series of peaks in $S_e$. However, as the chemical potential increases, the magnitude of the peaks reduces.
The reason is that the considered setup possesses four terminals that all exhibit an increasing number of propagating modes with increasing $\mu$. Thus, inter-mode scattering is more and more likely to happen. Then, the staircase behavior of $\Tsb$ diminishes and transforms into oscillations.

We close this section with a remark on the average sign of the Seebeck coefficient $S_e$. It is positive in the conduction band and negative in the valence band which reflects the sign of the corresponding excitations in a given band.

\section{Spin Nernst effect induced by bulk states}

A spatial dependence of model parameters, like, for instance, an in-plane electrostatic potential $V_0$ or the mass confinement potential $V_m$, can generate a transverse spin current resulting in a spin Hall signal for the metallic bulk states \cite{Rothe10}. This phenomenon has been previously analyzed in Refs. \cite{YokoyamaNagaosa09,Guigou11,Yamakage11} in the context of charge and spin transport properties at interfaces between metals and quantum spin Hall systems. As already mentioned above, the spin Hall conductance gives rise to the spin-Nernst  signal from the bulk states through the Mott-like relation \eqref{Mottlike}. Therefore, in the next two subsections we will focus on analytical models to describe the scaling of the spin current and spin Hall conductance with the band structure parameters and compare our intuitive analytical models with the numerics.

\begin{figure}
\includegraphics[scale=0.3]{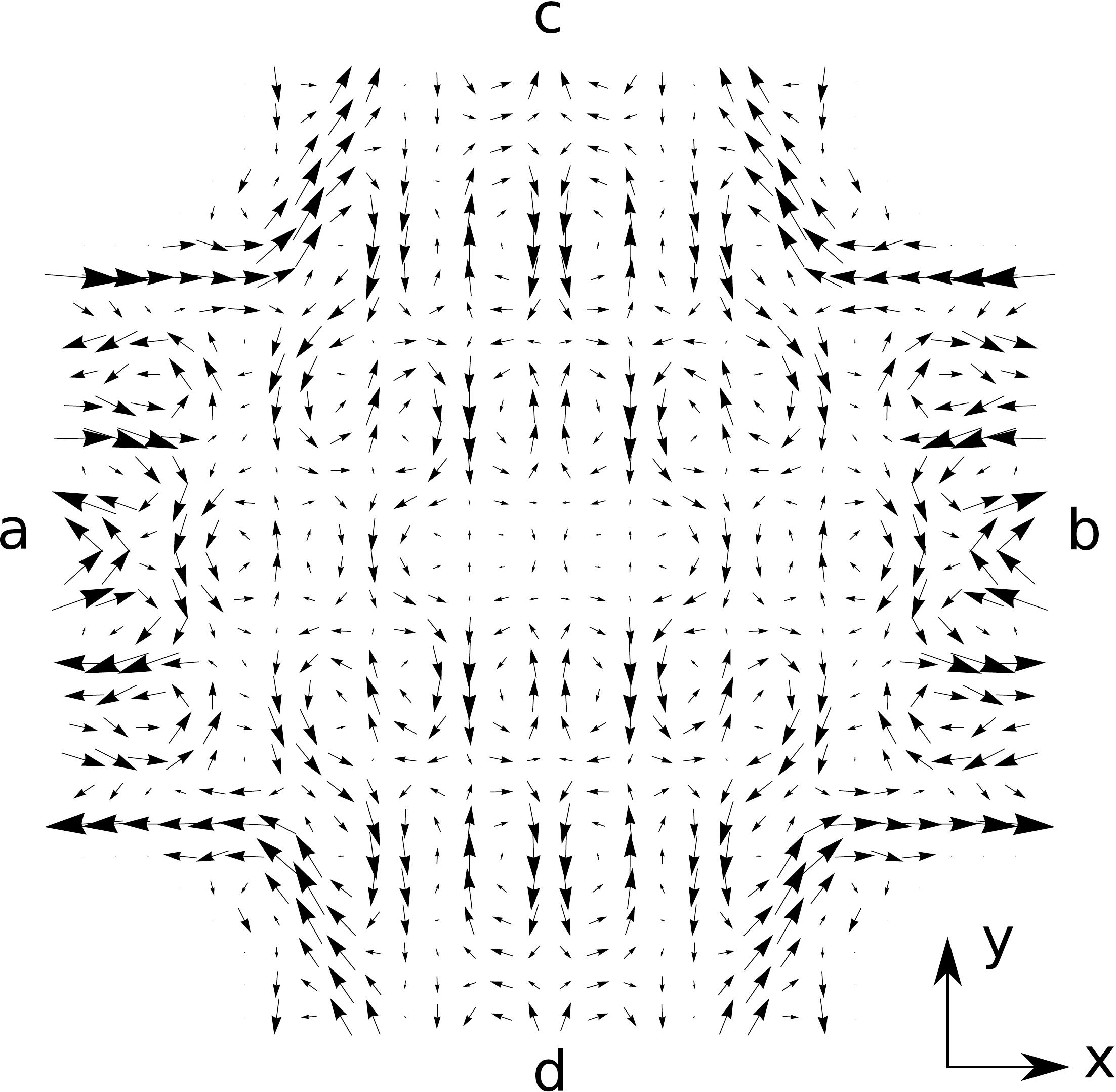}
\caption{\label{currentMap}
Local spin current $\vec{J}^z$ for normal metallic regime, and electrical bias, $\mu_{\mathsf{a}} - \mu = - (\mu_{\mathsf{b}} -\mu)$ with
$\mu = 0.4 t_0$ and $\Ttemp = 0$.
}
\end{figure}

First however, to visualize the formation of the spin Hall effect at the sample boundary, it is instructive to plot the local spin current density in the numerical 4-band model. In order to do so, we first
define a local spin current operator by
\begin{align}
\hat{\vec{J}}^z(\vec{r}') = \frac{1}{i} \{ [ \hat{\vec{r}}, H ],  \delta(\hat{\vec{r}}-\vec{r}') \} \sigma_z,
\end{align}
where $\sigma_z$ is a Pauli matrix that acts on the spin space of the 4-band model.
On the basis of the NEGF, it is then straightforward to evaluate the expectation value of the spin current operator at $\Ttemp=0$, which can be expressed as
\begin{align}
\vec{J}^z(\vec{r}) = \sum_p \Tr \left[ \hat{\vec{J}}^z(\vec{r}) G^R \Gamma_p G^A \right] \mu_p.
\end{align}
In Fig. \ref{currentMap}, the local spin current is shown for the normal metallic bulk regime and an electrical bias
applied from left to right. One clearly recognizes a spin current flowing along the edges.

Note that our model shows local spin currents already at equilibrium. However at equilibrium, the spin current integrated over the cross section of a lead cancels and thus does not enter the spin Hall signal. For clarity, the local spin current that we show in Fig. \ref{currentMap} is only the non-equilibrium part.

The rest of this section is organized as follows: parts A and B deliver two complementary approaches to explain the interface spin current that is transverse to 
 the potential gradient.
In part C, the connection between these interface spin currents and the spin-Nernst coefficient gives us a qualitative understanding of the behaviour of spin-thermo effects for the bulk states.

\subsection{Effective 2-band models}

Here, we show that an anomalous spin-dependent velocity naturally appears within effective 2-band spin-diagonal electron or hole band models obtained by perturbatively folding down the 4-band model (see Eq.~\eqref{H4bdmodel}). We apply third order quasi-degenerate perturbation theory similar to our previous work \cite{Rothe10}.

The diagonal part of the Hamiltonian is
$h_0 = \mathrm{diag}(M, -M)$.
For the perturbation part, we consider  $h'^{\up} = \left(\begin{array}{cc} \tilde{V}_e & A \hat{k}_+ \\ A \hat{k}_- & \tilde{V}_h \end{array}\right)$ and
$h'^\down = h'^{\up*}(-k)$, where $\tilde{V}_{e/h} = V_{e/h} \mp B k^2 - D k^2 $ and $V_{e/h} = V_0 \pm V_m$; the subscripts $e$ and $h$ refer to electron and heavy hole bands, respectively.
The $B$ and $D$ parameters will not enter  to the spin current in third order perturbation theory.
Note that, as compared to Eq. (\ref{H4bdmodel}), we allow for a finite in-plane potential $V_0$ in this analysis.
 Treating $\hat{k}_\pm$ as operators acting on a perturbing potential, we obtain the spin-dependent effective 2-band Hamiltonians as follows (showing only the third order)
\begin{align}
& h_{\rm eff,e}^\up = \frac{1}{8 M^2} \left( 2 A^2  \hat{k}_+ \tilde{V}_{h} \hat{k}_-  - \{ A^2 \hat{k}^2, \tilde{V}_{e} \} \right),
\\
& h_{\rm eff,h}^\up = \frac{1}{8 M^2} \left( 2 A^2  \hat{k}_- \tilde{V}_{e} \hat{k}_+ - \{ A^2 \hat{k}^2, \tilde{V}_{h} \} \right),
\end{align}
and $h_{\rm eff,e/h}^\down = (h_{\rm eff,e/h}^\up)^*$.
The lowest order spin-dependent term of the effective electron/heavy hole model is thus given by
\begin{align}
\nonumber & h_{\rm Pauli,e/h}  = \frac{1}{2} \left( h_{\rm eff,e/h}^\up - h_{\rm eff,e/h}^\down \right) \sigma_z
\\
 & =  -\frac{A^2}{4M^2} \left(\nabla (\pm V_0 - V_m) \times \vec{k}\right)_z \sigma_z.
\end{align}
In the Heisenberg picture, we obtain a spin-dependent anomalous velocity
\begin{eqnarray}
\vec{v}^{an,e/h} &=& \frac{1}{i \hbar} [ \vec{r}, h_{Pauli,e/h} ] \\
&=&  \frac{A^2}{4 M^2} \sigma_z \left(\begin{array}{c} - \partial_y\\ \partial_x\\ 0 \end{array}\right) (\pm V_0 - V_m). \nonumber
\end{eqnarray}
Since $\vec{v}^{an,e/h} \perp \nabla (V_0 \mp V_m)$, we expect to see a spin current
along the edge of the sample, similarly to the spin current carried by the edge states,
but now the effect is induced by the bulk modes.

Note that the assumptions for a valid perturbation theory are quite restrictive.
The condition  $A k \ll 2|M|$  restricts the energy range to about $|E| \lesssim 0.2 t_0$. Further, this approach works in the inverted regime only when one considers the bulk states and assumes a direct gap.
The main drawback of this perturbative approach is, however, that it assumes the variation of the potentials $V_0$, $V_m$ small compared to the gap $2 |M|$, which
is not the case for the numerical confinement potential. Therefore, although we expect to find qualitative results by this approach, it is important to compare it with the non-perturbative model including hard wall boundary conditions which will be done in the next subsection.

\subsection{Hard wall boundary spin current}

In this subsection, we present a complementary explanation of the spin current carried by the bulk states, valid also beyond the parameter regime $A k \ll 2|M|$, demonstrating that the reflection
of an incident wave at a hard wall boundary  leads to a spin current along the boundary. 
Due to a phase offset, this spin current persists even for a superposition of waves incident at different angles.
We will show below that in the regime $Ak \ll 2M$ the spin current scales like $A^2/M$.
Interestingly, we observe that the explanation of the spin current given here seems to be close to
what is seen in the numerical 4-band tight binding model, because the numerically calculated spin Hall effect indeed scales like $A^2/M$ in the parameter regime $Ak \ll 2M$ (with  $M>0$).

In the model we consider now, the hard wall boundary condition for the envelope function is given by $\psi(y=0) = 0$.
While the direction of the outgoing beam is restricted by the energy and momentum conservation laws and is not
spin-dependent, there is a spin-dependent phase shift between incident and reflected wave.
Remarkably, even for an incident wave normal to the interface, a spin current moving along the interface is generated. In case this interface is bent,
like it happens at the sample boundary connecting two perpendicular leads, it will transform into a spin Hall signal (like in Fig.~\eqref{currentMap}).

We start with the following ansatz for the spin $\uparrow$ wave function
\begin{equation}
\label{ansatzL}
\psi^\up(y) = \langle y | \psi^\up \rangle = e^{i k_y y} u_{k_y} + r e^{-i k_y y} u_{-k_y} + c e^{\lambda y} u_{i \lambda},
\end{equation}
where the plane wave dependency on $x$ has been separated off. $u_{\pm k_y}$ and $u_{i \lambda}$ denote the spinors for fixed energy $E$ and momentum $k_x$.
The condition $\psi^\up(0) = 0$ gives two equations for the coefficients $r$ and $c$.
The corresponding coefficients for spin down can be found by replacing $k_x \to -k_x$.
The operators of transverse velocity, $V^\up_x(k_x) = \frac{1}{\hbar} \frac{\partial h^\up}{\partial k_x}$ and
$V^\down_x(k_x) = -V^\up_x(-k_x)$ are independent of $k_y$ and complex-valued matrices.
In the following, we will plot both spin up (in blue) and spin down currents (in red), evaluated by
\begin{align}
 j^\up_{k_x}(y) = \bra{\psi^\up} \delta(y-\hat{y}) V_x^\up \ket{\psi^\up}, \; j^\down_{k_x}(y) = -j^\up_{-k_x}(y).
\end{align}
It is easy to see that the spin current
\begin{align}
 j^s(y) = j^\up_{k_x}(y) - j^\down_{k_x}(y) = j^\up_{k_x}(y) + j^\up_{-k_x}(y)
\end{align}
is symmetric in the angle of incidence $\theta = \tan^{-1} \frac{k_x}{k_y}$.

\begin{figure}
\includegraphics[scale=.7]{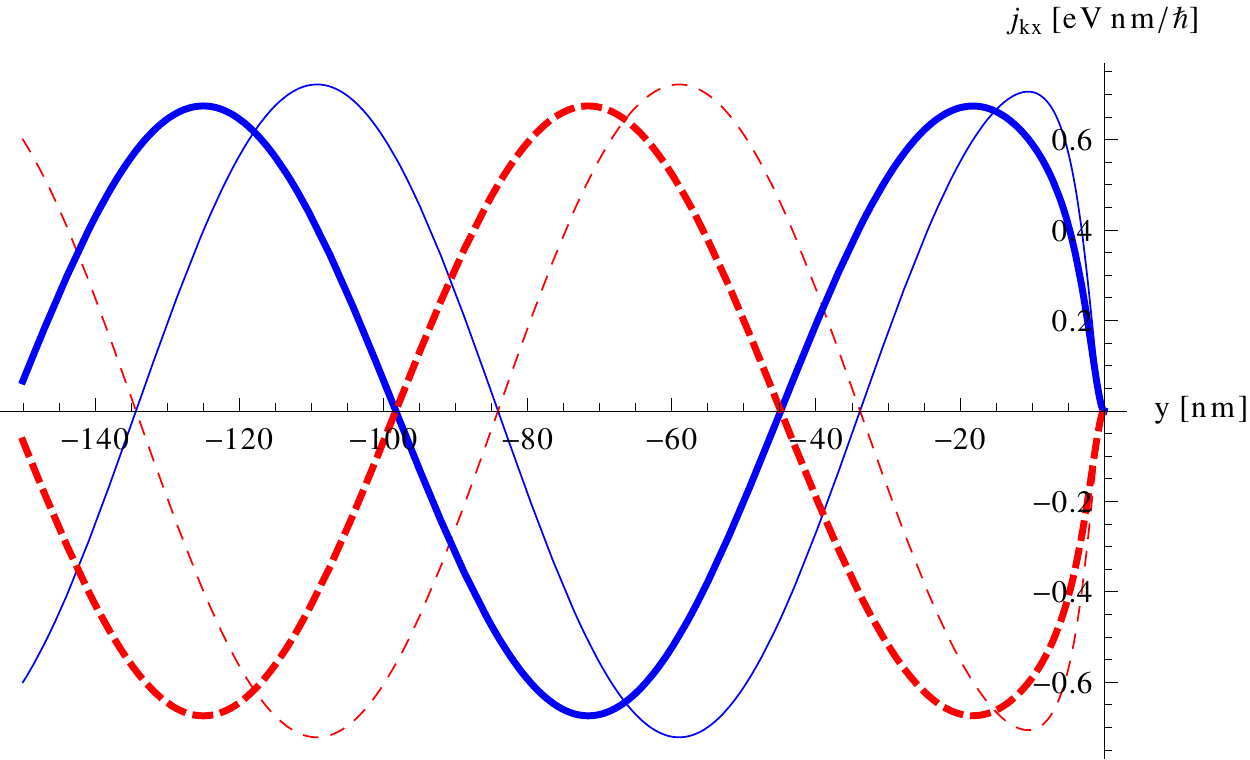}
\caption{
\label{currentxHardwall}
(Color online)
The spin-dependent currents parallel to the interface at $y=0$ are plotted, with $j^\up_{k_x}(y)$ in blue (solid) and $j^\down_{k_x}(y)$ in red (dashed).
For perpendicular angle of incidence ($k_x=0$), the phase difference $\Delta \phi$ is always $\pi$.
The energy is $E=0.3 t_0$ in all cases. The thick/thin lines show the normal/inverted regime with $M=0.1 t_0$ and $M=-0.1 t_0$,
respectively.
The other parameters are the same as for the lattice model.
}
\end{figure}

Figure \ref{currentxHardwall}  shows the spin up and down currents for typical parameters and the energy in the conduction band.
The superposition of incoming and reflected
propagating waves leads to an oscillating pattern.
We are interested in the phase shift between spin up and down.
The direct terms in $j^\up_{k_x}(y)$ (i.e. two incoming or two outgoing propagating modes) are constant in $y$ and
current conservation dictates that the incoming and reflected currents are the same.
Rotational invariance of the BHZ Hamiltonian and current conservation dictate that $|r^2| = 1$
independent of the spin.
Because of time reversal symmetry, the current of the direct terms is independent of the spin
and thus, the direct terms do not contribute to $j^s(y)$.

The interference term between the incoming and outgoing modes in $j^\up_{k_x}(y)$ is given by
\begin{align}
 \nonumber & 2 \mathrm{Re}\left[\bra{u_{k_y}} V_x \ket{u_{-k_y}} r e^{-2i k_y y} \right]
 \\
 & = \left| \bra{u_{k_y}} V_x \ket{u_{-k_y}}  r \right| 2 \cos(2 k_y y - \phi_1^\up - \phi_2^\up),
\end{align}
where $\phi_1^\up  = \arg(\bra{u_{k_y}} V_x \ket{u_{-k_y}})$ and  $\phi_2^\up = \arg(r)$.
In Ref. \cite{YokoyamaNagaosa09}, $\phi_2^\up -\phi_2^\down$ is called the angle of giant spin rotation.
At $k_x = 0$, we have
\begin{align}
 \left.\bra{u_{k_y}} V_x \ket{u_{-k_y}}\right|_{k_x=0} = \frac{i A^2 k_y}{\sqrt{A^2 k_y^2+(M-Bk_y^2)^2}},
\end{align}
where $k_y$ is fixed by the energy. A first-order expansion in $k$, valid in the regime $A k \ll 2|M|$ yields
\begin{align}
\label{jsy}
 j_s(y) \propto \frac{A^2 k}{|M|}.
\end{align}
In contrast to \cite{Guigou11}, the spin current in our  analytical analysis  is connected only with the propagating solutions as explained above. As one can see from Fig.~ \ref{currentxHardwall}, where the evanescent modes are included, the periodicity of $j^\up_{k_x}(y)$ and $j^\down_{k_x}(y)$ is only slightly affected which means that the evanescent contribution at least for the normal regime is minor and Eq.~\eqref{jsy} still holds.

\begin{figure}
a) \includegraphics[scale=1]{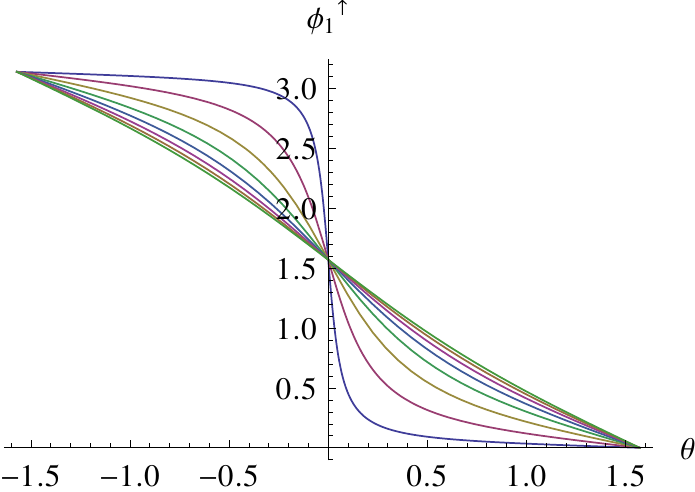}
\\
b) \includegraphics[scale=1]{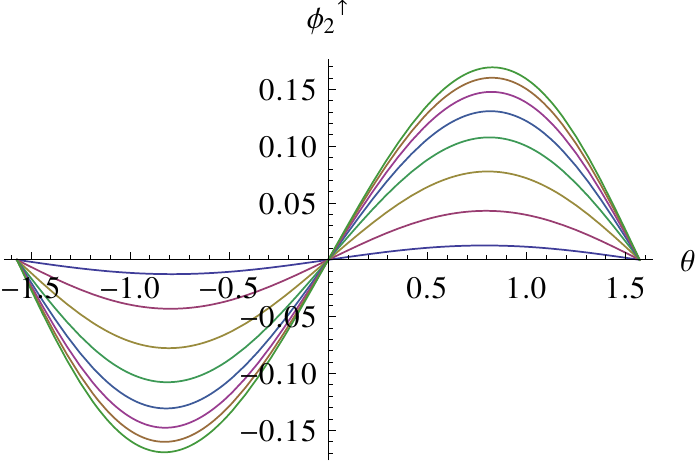}
\caption{
\label{phioftheta}
(Color online) The spin up phase shift of the reflected current is $\phi_1^\up + \phi_2^\up$.
a)  The phase (of the velocity matrix element) $\phi_1^\up$ as a function of the angle of incidence. It has the symmetry
$\phi_1^\up(-\theta) = \pi - \phi_1^\up(\theta)$. Different colors (blue,red,orange, ...) correspond to different choices of $A=0.05, 0.1, ...,0.4 \text{eV nm}$, where
the limit $A\to 0$ gives a step function.
Parameter values are $E = 0.3 t_0$, $M = 0.1 t_0$ and $B,D$ have values as in Section IIA.
The picture does not change qualitatively, if we change the parameters of the underlying model.
b)
The phase (of the reflection coefficient) $\phi_2^\up$ for the same parameters.
For low values of $A$, it is proportional to $A^2$, whereas for large values of $A$ it saturates.
}
\end{figure}
Let us now analyze the phase relations between spin up and spin down currents  more closely.
The two phases $\phi_1^\up$ and $\phi_2^\up$ (defined above) behave differently as a function of the angle of incidence, as shown in Fig. \ref{phioftheta}.
We find the symmetries $\phi_1^\up(\theta) = \pi-\phi_1^\up(-\theta)$ and $\phi_2^\up(\theta) = -\phi_2^\up(-\theta)$.
For $A \to 0$, $\phi^\up(\theta) = \phi_1^\up(\theta) + \phi_2^\up(\theta)$ becomes a step function, with $\phi^\up(0) = \pi/2$.
We are interested in $\Delta \phi = \phi^\up - \phi^\down$.
For this, we again use a symmetry.
If we flip the spin,
$r^\down(k_x) = r^\up(-k_x)$ implies that $\phi_2^\down(-\theta) = \phi_2^\up(\theta)$ and
$V^\down(k_x) = -V^\up(-k_x)$ implies that $\phi_1^\down(-\theta) = \pi + \phi_1^\up(\theta)$. Thus,
\begin{align}
\Delta \phi = \phi_1^\up + \phi_2^\up  - \phi_1^\down - \phi_2^\down = 2 (\phi_1^\up + \phi_2^\up)
\end{align}
with $\Delta \phi(\theta) = 2 \pi - \Delta \phi(-\theta)$.
For not too small parameters $A$ and small $\theta$, the constant phase shift $\Delta \phi(0) = \pi$ is dominant.
This phase shift ensures that the sign of the spin current  is well-defined over
a large range of $\theta$.
Therefore, even the superposition of many incident modes at different angles (not included in this simple analysis) would lead to a well-defined sign of the spin current near the interface, 
while far from the interface, the spin current will be suppressed by the oscillations.

\subsection{Spin-Nernst signal for the bulk metallic regime}
\begin{figure}
(a)\includegraphics[scale=.3]{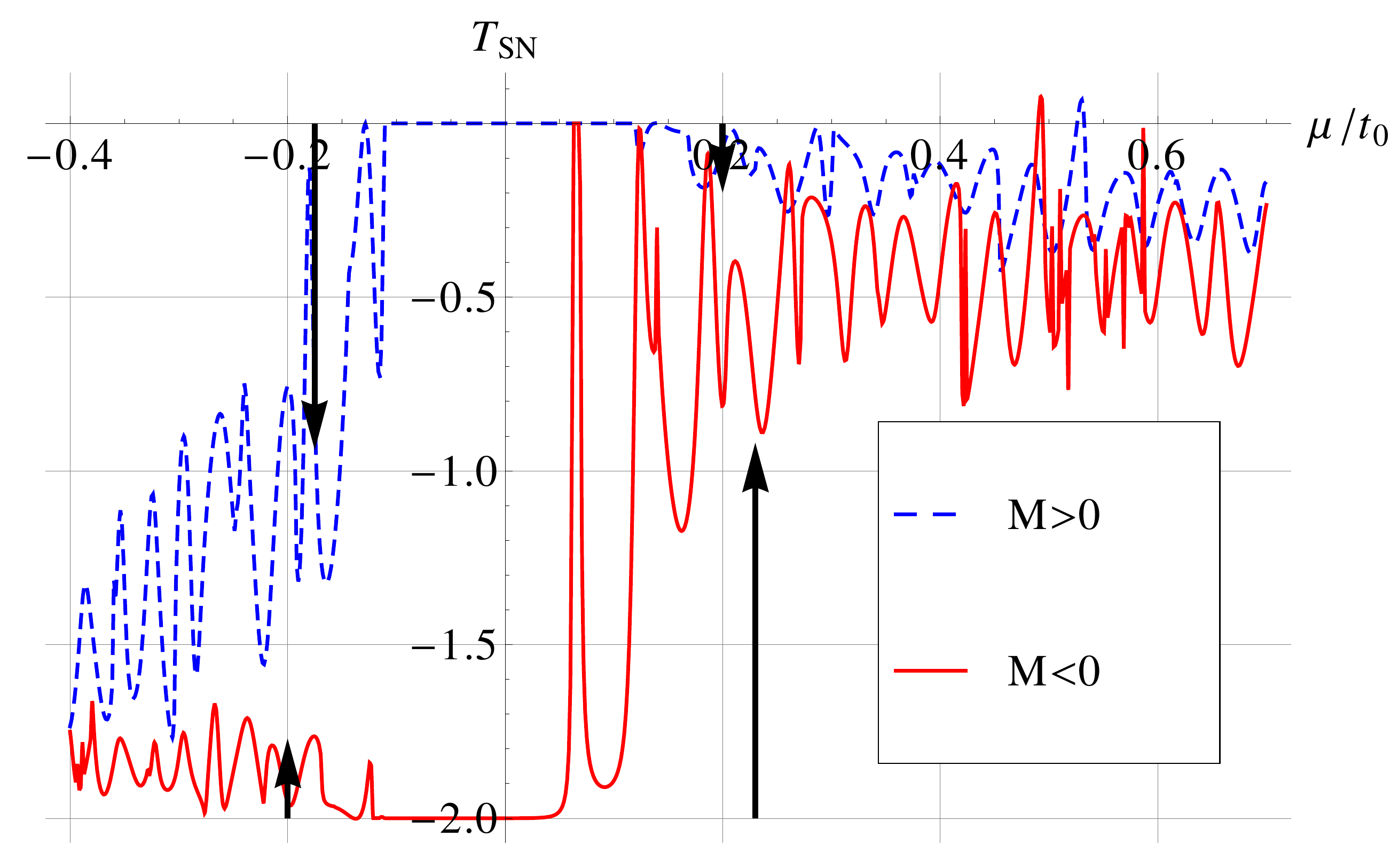}
\\
(b)\includegraphics[scale=.34]{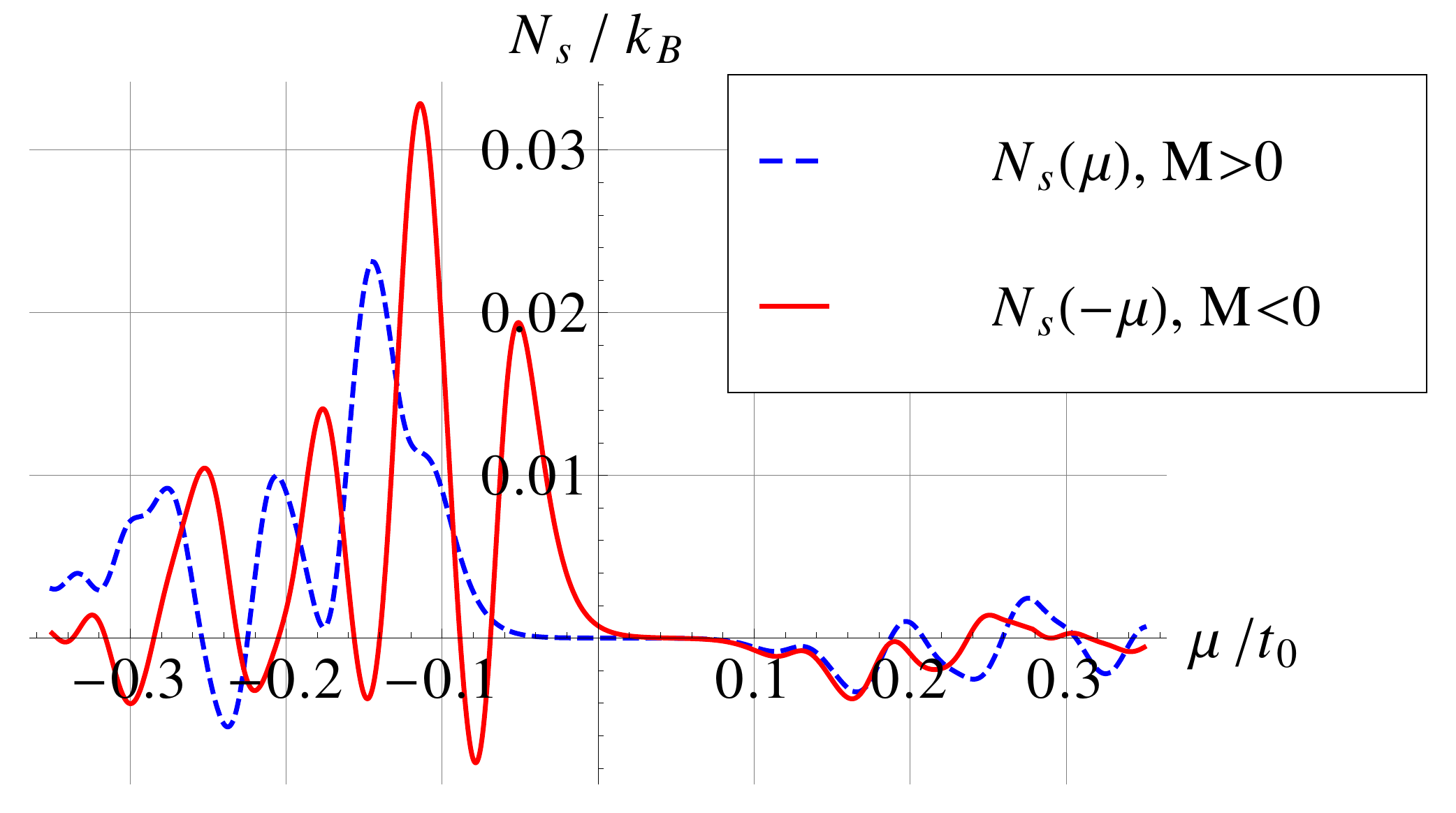}
\caption{\label{snernstBulk} (Color online)
(a) Spin Nernst transmission function $\Tns(\mu)$ for the normal regime ($M = 0.1 t_0$, blue dashed) and
inverted regime ($M = -0.1 t_0$, red solid). In the bulk gap, the edge states give rise to a quantized spin Hall
conductance of $e/2\pi$, except for the mini-gap.
The black arrows are scaled according to the expected factor of $\frac{-m_h}{m_e} \approx 4.7$
between conduction and valence band spin Hall effect (see text).
(b) Spin Nernst effect at $\Ttemp =  \mathrm{5~K}$, for normal (blue dashed) and inverted regime (red solid).
The latter is flipped horizontally, so we can compare signals of the same band character.
}
\end{figure}

In the preceding sections, we showed that the spin current can be understood by an anomalous velocity or a spin-dependent phase shift.
The expressions we have obtained do not depend on the effective band mass  (considering the lowest order in $A k/M$).
We will now show that such a scaling of the spin current leads to  a Nernst signal proportional  to the effective band mass.

Let us assume that the applied difference in the chemical potential $\Delta \mu$ generates the spin and the charge responses in the system.
Then $G_{sH} = \frac{I_s}{I} G_{xx}$. For a given number of modes $G_{xx}$  is approximately constant.
Therefore, using Eq.~\eqref{jsy}, one can see that $G_{sH} \sim m_{e/h} A^2 /M$, where $m_{e/h}$ is the mass of the electron/ heavy hole band, respectively.
To the lowest order in $Ak/M$, the effective 2-band and 4-band masses coincide and the 4-band effective masses $m_{e/h}$ are given by  $\hbar^2/2m_e=-D-B$ and $\hbar^2/2m_h=-D+B$.
Correspondingly, through the relation between the spin Hall conductance and the spin Nernst  transmission signal $ \Tns \sim G_{sH} \sim m_{e/h}$. 
The last dependence can be easily seen in the limit 
for $D=0$,  when the band structure of the BHZ model is particle-hole symmetric.
Then, $m_e = -m_h$ which is consistent with $\Tns \sim m_{e/h}$ and the symmetry 
relation  $\Tns(\mu) = \Tns(-\mu)$ in that case.
Figure \ref{snernstBulk} (a)  shows numerical results for $\Tns$ as a function of the chemical potential.
 In the 4-band model, the ratio of valence and conduction band effective masses is $\frac{-m_h}{m_e} \approx 4.7$.
The black arrows are drawn for comparison of $\Tns$ in conduction and valence band and are scaled by the factor $-m_h/m_e$.
Their position is chosen for energies corresponding to 4 propagating modes in the leads (counting spin), not
counting edge states. For the normal regime (see dashed lines in Fig.~\ref{snernstBulk} (a)) the scaling of the numerical $\Tns$  is very close to what we predicted from the analytical approaches.
In Fig. \ref{snernstBulk}(b), we show the corresponding spin Nernst signal. In the normal regime,
we qualitatively find $N_s \propto |m_i|$ ($i=e,h$) as expected from the Mott-like relation in combination with Fig. \ref{snernstBulk}(a).

In the inverted regime (solid red line) we must consider that near the bulk gap,
the band character (E/H) has changed (compare the red/blue coloring in Fig. \ref{dispNINV}); therefore, the band for $\mu > 0$ gets a heavy hole character.
Further, as long as the edge states do not yet merge to the bulk, they are responsible for an offset
of $\Tns = -2$. The black arrows again indicate the factor $\frac{-m_h}{m_e}$ that we expect for the comparison of
conduction and valence band signals at the same number of contributing modes, however now we are measuring the signal  from the level of the edge states.
Analyzing numerically the scattering matrix we find that in the valence band the contribution to $\Tns$ of  bulk and edge states are additive, while this is not the case for the conduction band.
Taking into account this fact, it is suprising that the simple analytical analysis  applicable to the normal regime still describes qualitatively the numerics.
 We believe that  this might be the case, because  the first  bulk state resembles the edge state character, and our argument  about the symmetry of $\Tns(\mu) =\Tns(-\mu)$ for the particle-hole symmetric Hamiltonian still holds.

\section{Conclusion}

We have analyzed the thermoelectric transport in four-terminal setups of HgTe/CdTe quantum wells with a particular emphasis on spin-dependent effects due to spin-orbit coupling. Thereby, we have used a combination of analytical and numerical methods to analyze spin-dependent transport phenomena. The Seebeck and the spin Nernst signal show a peculiar dependence on the parameters of the Bernevig-Hughes-Zhang model which can be qualitatively understood as originating from a spin Hall effect that arises at in-plane potential or confinement boundaries of the system. We have demonstrated that the spin Nernst effect is a strong experimental tool to get a better understanding of the mini-gaps that arise due to the spatial overlap of edge states on opposite sample boundaries. Most interestingly, we have derived a Mott-like relation between the spin Nernst coefficient and a smoothed spin Nernst transmission function that is valid to all orders in the temperature difference between the warm and the cold reservoir. Our findings might help to optimize future experiments on thermoelectric transport properties of two-dimensional topological insulators.

Financial support by the German Science Foundation (DFG, SPP 1285), the Helmholtz Foundation (VITI), and the European Science Foundation (ESF) is gratefully acknowledged.

\appendix

\section{Tight binding Hamiltonian}
\label{appendixTBModel}
Using the representation of the plane wave annihilation operator in the basis of lattice sites $(j,l)$,
$c_{\vec{k}} = \frac{4 \pi^2}{a^2} \sum_{j,l}  e^{i a (k_x j + k_y l)}  c_{j,l}$, we obtain the
following substitution rules for the continuum model momentum operators,
\begin{align}
\nonumber \hat{k}_x \to & \int_{BZ} d^2 k \, \frac{1}{a} \sin(a k_x) c^\dagger_{\vec{k}} c_{\vec{k}}
\\
& =  \frac{1}{2 i a} \sum_{j,l} c^\dagger_{j+1,l} c_{j,l} - c^\dagger_{j-1,l} c_ {j,l},
\end{align}
\begin{align}
\nonumber \hat{k}_x^2 \to &   \int_{BZ} d^2 k \, \frac{1}{a} \left(2 - 2 \cos(a k_x)\right) c^\dagger_{\vec{k}} c_{\vec{k}}
\\
& =  -\frac{1}{a^2} \sum_{j,l} c^\dagger_{j+1,l} c_{j,l} - 2 c^\dagger_{j,l} c_ {j,l} + c^\dagger_{j-1,l} c_ {j,l},
\end{align}
and analogous rules for $\hat{k}_y$ and $\hat{k}_y^2$.

We define matrices on the band space
$\Gamma^1 = \tau_x  \sigma_z$, $\Gamma^2 = -\tau_y \sigma_0$, $\Gamma^3  = \frac{\tau_x + \tau_z}{2} \sigma_y$,
$\Gamma^4 = -\frac{\tau_x + \tau_z}{2} \sigma_x$ and $\Gamma^5 = \tau_z \sigma_0$
where $\tau_i$ are the Pauli matrices acting on the E/H space and $\sigma_i$ acts on the spin space 
($\tau_0 , \sigma_0$ are unit matrices).
Then, the lattice Hamiltonian corresponding to $H$ of Eq. \eqref{H4bdmodel} reads
\begin{align}
 H_{\rm tb} = & \sum_{j,l} \Big[
 \left( \frac{A \Gamma^1 + R_0 \Gamma^3}{2 i a}  +  \frac{B \Gamma^5 + D \mathbf{1}}{a^2} \right)
 c^\dagger_{j+1,l} c_{j,l}
\nonumber \\ &
 + \left(-\frac{A \Gamma^1 + R_0 \Gamma^3}{2 i a}  +  \frac{B \Gamma^5 + D \mathbf{1}}{a^2}  \right)
  c^\dagger_{j-1,l} c_ {j,l}
\nonumber \\ &
 + \left(\frac{A \Gamma^2 + R_0  \Gamma^4 }{2 i a}  + \frac{B \Gamma^5 + D \mathbf{1}}{a^2} \right)
  c^\dagger_{j,l+1} c_{j,l}
\nonumber \\ &
 + \left(-\frac{A \Gamma^2 + R_0 \Gamma^4}{2 i a}  +  \frac{B \Gamma^5 + D \mathbf{1}}{a^2} \right)
   c^\dagger_{j,l-1} c_ {j,l}
\nonumber \\ &
-  \left(\frac{4}{a^2} (B \Gamma^5 + D \mathbf{1})  + M \Gamma^5 \right)
  c^\dagger_{j,l} c_ {j,l}  \Big] ,
\end{align} 
where the summation over the grid points $(j,l)$ is restricted by the geometry of the sample and we only included the most important linear Rashba term,
 proportional to $R_0$ in Eq.~(\ref{hra}).

\section{Mott-like relation}
\label{appendixMottlike}
In this Appendix, we show how the Mott-like relation of Eq. \eqref{Mottlike}
can be generalized to finite temperatures.
For this we consider the Fourier transform of transmission functions
$\Delta T_{\mathsf{c},q}(\tau) = \frac{1}{2\pi} \int d E e^{-i E \tau} \Delta T_{\mathsf{c},q}$.
The spin Nernst effect is defined as
\begin{equation}
 N_s = \frac{I^s_\mathsf{c}}{2 \Delta \Ttemp} = \frac{1}{8 \pi \Delta \Ttemp} \int d E  \sum_q \Delta T_{\mathsf{c},q}(E) (f_\mathsf{c} - f_q),
\end{equation}
where the potential $\mu$ is assumed to be the same for all leads, while the temperatures may differ.
The integral has the form of a convolution. The Fourier representation is
\begin{align}
N_s = \frac{1}{4\pi} \int_{-\infty}^\infty d \tau \frac{i}{\tau} e^{-i \mu \tau} \sum_q \frac{\Delta T_{\mathsf{c},q}(\tau)}{\Delta \Ttemp}
\left.\frac{x}{\sinh x}\right|_{x=\pi \tau k_B \Ttemp_q}^{x=\pi \tau k_B \Ttemp_\mathsf{c}}.
\end{align}
Further, we define a symmetric ``smoothing'' function that depends on temperatures of the leads $\mathsf{c}$ and $q$ as
\begin{align}
 F^q(\tau) =  \frac{3}{\pi^2 k_B^2 \Ttemp_\mathsf{c} \Delta \Ttemp}  \left.\frac{x}{\sinh x}\right|_{x=\pi \tau k_B \Ttemp_q}^{x=\pi \tau k_B \Ttemp_\mathsf{c}}.
\end{align}
If we put $\Ttemp_\mathsf{c} - \Ttemp_q = \Delta \Ttemp$, we find with  $\Delta \Ttemp \to 0$
\begin{align}
\label{Fqlin}
 F^q(\tau) \approx  3 \left.\frac{x\coth x - 1}{x \sinh x}\right|_{x=\pi \tau k_B \Ttemp_\mathsf{c}},
\end{align}
which has a width of $\Delta \tau \approx \frac{4}{\pi k_B \Ttemp_\mathsf{c}}$
and $F^q(0) \to 1$.
Now, we define a temperature-smoothed spin Nernst transmission function as
\begin{equation} \label{B5}
\Tnssm(\tau) = \sum_q F^q(\tau) \Delta T_{\mathsf{c},q}(\tau),
\end{equation}
which implies that $\Tnssm(E) = \int d\tau e^{-i E \tau} \Tnssm(\tau)$ is real.
Finally, we obtain the relation
\begin{equation}
\label{MottlikeExact}
 N_s(\mu)
 = \frac{\pi k_B^2 \Ttemp_\mathsf{c}}{12} \left.\frac{d \Tnssm(E)}{d E}\right|_{E=\mu},
\end{equation}
which is exact to all orders in $\Ttemp_\mathsf{c}$ and $\Delta \Ttemp$.
The meaning of the latter equation is the following one: First taking the 
derivative $\partial_E \Tns$ and then smoothing by temperature
is the same as first smoothing with a modified smoothing kernel and then taking the derivative.
\newline ~

\bibliographystyle{apsrev}
\bibliography{notes}

\end{document}